\documentclass[12pt]{iopart}

\usepackage{graphicx}
\begin{document}

\title{Finite-size effects from higher conservation laws for the one-dimensional Bose gas}

\author{Erik Eriksson$^1$ and Vladimir Korepin$^2$}
\address{$^1$ Department of Physics, University of Gothenburg, SE 412 96 Gothenburg,
Sweden \\
$^2$ C. N. Yang Institute for Theoretical Physics, State University of New York at Stony Brook, NY 11794-3840, USA}
\ead{erik.eriksson@physics.gu.se and korepin@insti.physics.sunysb.edu}

\begin{abstract}
We consider a generalized Lieb-Liniger model, describing a one-dimensional Bose gas with all its conservation laws appearing in the density matrix. This will be the case for the generalized Gibbs ensemble, or when the conserved charges are added to the Hamiltonian. The finite-size corrections are calculated for the energy spectrum. Large-distance asymptotics of correlation functions are then determined using methods from conformal field theory.
\end{abstract}

\pacs{02.30.Ik, 05.30.Jp }


\section{Introduction}

There has recently been much interest in so called generalized Gibbs ensembles (GGEs), where the density matrix of a system is given by
\begin{equation} \label{gge}
\hat{\rho}_{GGE} = Z_{GGE}^{-1} \ e^{-\sum_n \beta_n \hat{Q}_n},
\end{equation}
generalizing the usual Gibbs ensemble for generic (non-integrable) systems where only the Hamiltonian and particle number operator appear in the exponent. Instead, for an integrable system, $\{ \hat{Q}_n \}$ is the complete set of local conserved charges, with
$\{ \beta_n \}$ the generalized inverse temperatures (Lagrange multipliers) and $Z_{GGE} = \textrm{Tr} \ e^{-\sum_n \beta_n \hat{Q}_n}$ the generalized partition function~\cite{jaynes}. For such quantum ensembles, expectation values of observables are obtained as
\begin{equation}
\langle \hat{O} \rangle_{GGE} = \textrm{Tr} \  \hat{\rho}_{GGE}  \hat{O},
\end{equation}
whereas the time-evolution is governed by the usual Schr\" odinger Hamiltonian. 

Recent experimental advances with quantum non-equilibrium dynamics of cold atoms~\cite{kinoshita}, where the GGE was proposed~\cite{rigol} to describe local large-time behavior, has spurred a great interest in possible equilibrium ensembles for integrable systems (see e.g. Refs.~\cite{polkovnikov,refs,ce} and references therein, and in particular Refs.~\cite{ck,ksci,mc2,bck} for treatments of the one-dimensional Bose gas). But effects from higher conservation laws in integrable systems have also been studied longer ago in the context of competing interactions in spin chains~\cite{tsvelik,frahm}.

It is therefore a timely question to now ask in general what sort of effects one can anticipate when incorporating higher conservation laws into the density matrix for an integrable model. In this paper we study finite-size effects and obtain conformal dimensions for the one-dimensional Bose gas (Lieb-Liniger model), when governed by a density matrix of the form (\ref{gge}). In particular, using the exact Bethe Ansatz solution we calculate the finite-size corrections for the energy and momentum. Comparing this to the expressions given by conformal field theory, we are able to obtain the large-distance asymptotics of correlation functions.

\section{Lieb-Liniger model}

The Lieb-Liniger model describes a one-dimensional Bose gas with point-like interaction through the Hamiltonian
\begin{equation} \label{HLL}
\hat{H} = \int dx \left[ \partial_x \Psi^{\dagger}(x) \partial_x \Psi(x) + c  \Psi^{\dagger}(x) \Psi^{\dagger}(x)\Psi(x) \Psi(x)  \right],
\end{equation}
with coupling constant $c>0$ and Bose fields $\Psi$ with equal-time commutation relations $\left[ \Psi(x),\Psi^{\dagger}(y)\right] = \delta(x-y)$ and $\left[ \Psi(x),\Psi(y)\right]  = \left[ \Psi^{\dagger}(x),\Psi^{\dagger}(y)\right]  = 0$ . The model is solved through the Bethe Ansatz \cite{liebliniger, lieb,kbi}, yielding eigenstates $| \{ \lambda_j \} \rangle $ given in terms of rapidities $\lambda_j$ satisfying the Bethe equations
\begin{equation} \label{bethe}
e^{i \lambda_jL} = - \prod_{k=1}^{N} \frac{\lambda_j - \lambda_k + i c}{\lambda_j - \lambda_k - i c}, \qquad j=1,...,N,
\end{equation}
for a system of $N$ particles in a box of length $L$ with periodic boundary conditions. The Bethe equations (\ref{bethe}) can equivalently be written as
\begin{equation} \label{bethe2}
L\lambda_j + \sum_{k=1}^{N} \theta(\lambda_j - \lambda_k) = 2\pi n_j, \qquad j=1,...,N,
\end{equation}
with $n_j$ integer when $N$ odd and half-integer when $N$ even, and $\theta(\lambda) = i \ln\left[ (ic+\lambda )/(ic-\lambda)\right]$. For non-zero wave function, $n_j \neq n_k$ when $j \neq k$.

The conserved charges $\hat{Q}_n$ have eigenvalues $Q_n$ given by
\begin{equation}
Q_n = \sum_{j=1}^N \lambda_j^n,
\end{equation}
the three lowest being particle number $N=Q_0$, momentum $P=Q_1$, and energy $E=Q_2$. Explicit expressions for some of the higher conserved charges $\hat{Q}_n$ can be found in Ref.~\cite{davies}. Now let us consider a generalized Hamiltonian $\hat{H}_G$ where all the conserved charges $\hat{Q}_n$ have been added to the Lieb-Liniger Hamiltonian (\ref{HLL}),
\begin{equation} \label{hg}
\hat{H}_G = \hat{H} + \sum_{n \neq 2} b_n \hat{Q}_n
\end{equation}
with coefficients $b_n$. Extremizing the entropy gives a density matrix of the form (\ref{gge})
\begin{equation} \label{g}
\hat{\rho}_{G} = Z_{G}^{-1} \ e^{-\beta \hat{H}_G},
\end{equation}
with $\beta = 1/T$ the inverse temperature and $Z_G$ the generalized partition function. Hence we can analyze both the situation of a system described by a generalized Gibbs ensemble (\ref{gge}) as well as a system where the Schr\" odinger Hamiltonian itself is given by $\hat{H}_G$ in Eq.~(\ref{hg}).

The eigenvalues $E(\{ \lambda_j\})$ of the generalized Hamiltonian $\hat{H}_G$ are given by
\begin{equation} \label{eg}
E(\{ \lambda_j\}) =  \sum_{j=1}^N \varepsilon_0 (\lambda_j)
\end{equation}
where the bare one-particle energy $\varepsilon_0 (\lambda)$ is given by the polynomial function
\begin{equation} \label{e0}
 \varepsilon_0 (\lambda) = \sum_{n=0}^{\infty} b_n \lambda^n
\end{equation}
with $b_2 = 1$, and $b_0=-h$ the chemical potential. It was shown in Ref.~\cite{mossel} that the generalized Yang-Yang thermodynamic Bethe Ansatz (TBA) equation becomes
\begin{equation} \label{yy}
\varepsilon(\lambda) + \frac{1}{2\pi \beta} \int_{-\infty}^{\infty} d\mu \, K(\lambda,\mu) \ln (1+ e^{-\beta \varepsilon(\mu)}) = \varepsilon_0 (\lambda),
\end{equation}
and that a solution exists provided that $\varepsilon_0 (\lambda)$ is bounded from below and $\lim_{\lambda \to \pm \infty}  \varepsilon_0 (\lambda) = + \infty$. The kernel $K(\lambda,\mu)$ is given by
\begin{equation}  \label{k}
 K(\lambda,\mu)= \theta'(\lambda-\mu) = 2c/(c^2 + (\lambda-\mu)^2),
 \end{equation}
and the equilibrium particle distribution function $\rho(\lambda)$ by
\begin{equation} \label{rho}
\rho(\lambda) = \vartheta(\lambda) \left( \frac{1}{2\pi} + \frac{1}{2\pi} \int_{-\infty}^{\infty} d\mu \, K(\lambda,\mu) \rho(\mu) \right)
\end{equation}
with the Fermi weight
\begin{equation} \label{theta}
\vartheta(\lambda) = \frac{1}{1+e^{\beta \varepsilon(\lambda)}}.
\end{equation}
This can also be written as $\rho(\lambda) = \vartheta(\lambda)\rho_t (\lambda)$, where
\begin{equation} \label{rhot}
\rho_t (\lambda) - \frac{1}{2\pi} \int_{-\infty}^{\infty} d\mu \,\vartheta(\mu) K(\lambda,\mu) \rho_t(\mu) = \frac{1}{2\pi}
\end{equation}
is the density of states. Similarly, the so-called dressed charge $Z(\lambda)$ is defined by
\begin{equation} \label{Z}
Z(\lambda) - \frac{1}{2\pi} \int_{-\infty}^{\infty} d\mu \,\vartheta(\mu) K(\lambda,\mu) Z (\mu) = 1,
\end{equation}
reflecting that for the Lieb-Liniger model $Z(\lambda) = 2\pi \rho_t (\lambda)$.

\section{Finite-size corrections at zero temperature}

\subsection{Energy}

Let us now investigate the finite-size corrections to the generalized energy (\ref{eg}), focusing on the zero-temperature limit. Then the set of numbers $n_j$ in Eq.~(\ref{bethe2}) that minimizes the generalized energy (\ref{eg}) is such that all $\lambda_j$ with $\varepsilon(\lambda_j)<0$ are occupied and the rest empty of particles. In the thermodynamic limit, this gives Fermi points $q_i^{\pm}$ wherever $\varepsilon(q_i^{\pm})=0$, with a filled Fermi sea $i$ for $q_i^{-} \leq \lambda \leq q_i^+$. The ground state energy in the thermodynamic limit can then be written as
\begin{equation} \label{eg0}
E_0 =  L \int_{-\infty}^{\infty} d\lambda \, \rho(\lambda) \varepsilon_0 (\lambda) = L \sum_i \int_{q_i^-}^{q_i^+} d\lambda \, \rho_t(\lambda) \varepsilon_0(\lambda),
\end{equation}
with finite-size corrections coming from the replacement of the sum in Eq.~(\ref{eg}) with an integral. These are obtained using the Euler-Maclaurin formula in complete analogy with the usual case~\cite{we,bm}, yielding for the leading corrections
\begin{equation} \label{eg0L}
E_0 =  L \sum_i \int_{q_i^-}^{q_i^+} d\lambda \, \rho_t(\lambda) \varepsilon_0(\lambda)    - \frac{\pi }{12 L}  \sum_{i,\nu}  | v_i^{\nu}|   ,
\end{equation}
where $i$ is the Fermi sea index and $\nu=\pm$ the right/left index. Here $v_i^{\pm}$ is the Fermi velocity at the Fermi point $q_i^{\pm}$, given by
\begin{equation}
v_i^{\pm} = \frac{1}{2\pi \rho_t(q_i^{\pm})} \frac{\partial \varepsilon}{\partial \lambda} \Bigg|_{\lambda=q_i^{\pm}}.
\end{equation}

Now we investigate the finite-size corrections of low-energy excited states, using the techniques found in Refs.~\cite{1979,bir,wet,w,hubbard,zvyagin}. First, we expand the energy $E(\{ q_i^{\pm}\}) = E_0 + \delta E $ to second order around the ground-state energy $E_0$ in Eq.~(\ref{eg0L}),
\begin{equation} \label{dnt3}
\delta E =  \sum_{i,\nu} \left( \frac{\partial E}{\partial  q^{\nu}_i } \right) \delta q^{\nu}_i  + \frac{1}{2}  \sum_{i,j,\nu,\nu'}   \left( \frac{\partial^2 E}{\partial  q^{\nu}_i \partial  q^{\nu'}_j } \right) \delta q^{\nu}_i  \delta q^{\nu'}_j
\end{equation}
with $\delta q^{\pm}_i $ the change in Fermi momentum $ q^{\pm}_i$ with respect to the ground state.
Now, 
\begin{equation} 
\hspace{-1cm} \frac{\partial E}{\partial  q^{\pm}_i } = \pm   L \rho_t ( q^{\pm}_i) \varepsilon_0 ( q^{\pm}_i)   +  L \sum_j \int_{q_j^-}^{q_j^+} d\lambda \, \frac{ \partial \rho_t(\lambda)}{\partial q^{\pm}_i  } \varepsilon_0(\lambda) =   \pm    L \rho_t ( q^{\pm}_i) \varepsilon( q^{\pm}_i)
\end{equation}
where $\varepsilon$ is the dressed energy (\ref{yy}). Since the ground state minimizes $E$, one has
\begin{equation} 
 \varepsilon( q^{\pm}_i) = 0 .
\end{equation}
Then Eq.~(\ref{dnt3}) becomes
\begin{equation} \label{dnt4}
\delta E =   \frac{L}{2}  \sum_{i,j,\nu,\nu'} \nu \delta q^{\nu}_i  \delta q^{\nu'}_j   \frac{\partial }{ \partial  q^{\nu'}_j }  \left[   \rho_t ( q^{\nu}_i) \varepsilon( q^{\nu}_i)  \right]   
\end{equation}
which is given by
\begin{equation} \label{dnt5}
 \delta E =   \frac{L}{2}  \sum_{i,\nu}  \nu    \rho_t ( q^{\nu}_i) \varepsilon'( q^{\nu}_i)  \left( \delta q^{\nu}_i  \right)^2  =  \pi L   \sum_{i,\nu}  |v_i^{\nu}|  \left[ \, \rho_t ( q^{\nu}_i)  \delta q^{\nu}_i  \,\right]^2
\end{equation}

Now we will express $\delta q^{\pm}_i $ in terms of the quantum numbers $\Delta N_i$ and $\Delta D_i$, defined as the changes in
\begin{eqnarray}
N_i = L \int_{q_i^-}^{q_i^+} d\mu \, \rho_t (\mu), \\
D_i = L\left( \int_{-\infty}^{q_i^-} - \int_{q_i^+}^{\infty}  \right) d\mu\, \rho_t (\mu) ,
\end{eqnarray}
compared to the ground state. Then
\begin{equation}  \label{dq}
\delta q^{\pm}_i = \sum_j \frac{\partial q^{\pm}_i}{\partial N_j} \Delta N_j + \sum_j \frac{\partial q^{\pm}_i}{\partial D_j} \Delta D_j.
\end{equation}
The Jacobian is found from
\begin{eqnarray}
\frac{\partial N_j}{\partial q^{\pm}_i} = \pm L \rho_t (q^{\pm}_i) \left[ \delta_{ij} \pm \int_{q^{-}_j}^{q_j^+} d\mu \, g(\mu|q^{\pm}_i) \right] , \label{dndq} \\
\frac{\partial D_j}{\partial q^{\pm}_i} =  L \rho_t (q^{\pm}_i) \left[ \delta_{ij} + \left( \int_{-\infty}^{q_j^-} - \int_{q_j^+}^{\infty}  \right)  d\mu \, g(\mu|q^{\pm}_i) \right] , \label{dddq}
\end{eqnarray}
where the function $g(\lambda | q^{\pm}_i)$ is defined through
\begin{equation} \label{gq}
g(\lambda|q^{\pm}_i) - \frac{1}{2\pi} \sum_k  \int_{q_k^-}^{q_k^+} d\mu \, K(\lambda,\mu) g(\mu|q^{\pm}_i) = \pm \frac{1}{2\pi} K(\lambda, q_i^{\pm} ).
\end{equation}
Eq.~(\ref{gq}) is equivalent to $g(\lambda|q^{\pm}_i) = \left[ \partial \rho_t (\lambda) / \partial q_i^{\pm}\right]  / \rho_t (q_i^{\pm})$. For the general case we need to consider the vectors
\begin{equation}
\vec{n} = \left( \begin{array}{c}
 N_1 \\ \vdots \\  D_1 \\ \vdots   
\end{array} \right) ,  \qquad \vec{p}  = \left( \begin{array}{c}
 q_1^- \\   q_1^+  \\   q_2^-  \\  \vdots 
\end{array} \right)
\end{equation} 
so that we can write
\begin{equation}
\Delta n_i = L \sum_j M_{ij}  \rho_t (p_j)  \delta p_j,  \qquad M_{ij} = \frac{1}{L \rho_t (p_j) }\frac{\partial n_i }{\partial p_j} ,
\end{equation}
with $\partial n_i  / \partial p_j$ given by Eqs.~(\ref{dndq})-(\ref{dddq}), and hence obtain $\rho_t ( q^{\pm}_i)  \delta q^{\pm}_i $ from
\begin{equation}
\rho_t(p_i) \delta p_i = \frac{1}{L} \sum_j (M^{-1})_{ij} \Delta n_j .
\end{equation}
Inserting this into Eq.~(\ref{dnt5}), and including the numbers $N_i^-$ and $N_i^+$ of particle-hole excitations at $q_i^-$ and $q_i^+$ respectively, gives the general expression
\begin{equation} \label{fsc}
   \delta E = \frac{\pi}{L} \sum_{i,\nu}  | v_i^{\nu} | \left[ 2N_i^{\nu}  + \left( \sum_j (M^{-1})_{q_i^{\nu}j} \Delta n_j   \right)^2 \right]  .
\end{equation}

This expression is simplified when the model has parity symmetry, e.g. around $\lambda=0$ so that only even powers of $\lambda$ appears in Eq.~(\ref{e0}) and hence $ \varepsilon_0(\lambda) = \varepsilon_0(-\lambda)$. If this is the case, we can consider the new pairs of Fermi points $q_j^{\pm} = \pm q_j$ ($q_1 > q_2 > ... > 0$), with a sea $j$ of either particles or holes between them, and Fermi velocities $|v_i^{\pm}| = v_i$. Now define
\begin{eqnarray}
N_j = L \int_{-q_j}^{q_j} d\mu \, \rho_t (\mu), \\
D_j = L  \left( \int_{-\infty}^{-q_j} - \int_{q_j}^{\infty}  \right) d\mu\, \rho_t (\mu) ,
\end{eqnarray}
i.e. $\Delta N_j$ being the number of particles/holes added to Fermi sea $j$ of particles/holes, and $d_j=\Delta D_j/2$ the number jumping from $q_j$ to $-q_j$. Let us now write 
\begin{eqnarray}
\hspace{-1cm} \frac{\partial N_i}{\partial q_j^{\pm}} =  \pm L \rho_t(q_j) \,2 \left[ \delta_{ij} - (-1)^j \int_{-q_i}^{q_i} d\mu \, g(\mu|q_j) \right] \equiv \pm L \rho_t(q_j) 2 Z_{ij} , \label{xmat} \\
 \hspace{-1cm}  \frac{\partial D_i}{\partial q_j^{\pm}} =  L \rho_t(q_j) \,2 \left[ \delta_{ij} - (-1)^j \left( \int_{-\infty}^{-q_i} - \int_{q_i}^{\infty}  \right)  d\mu \, g(\mu|q_j) \right] \equiv  L \rho_t(q_j) 2 Y_{ij} , \label{zmat}
\end{eqnarray}
with $g(\lambda | q^{\pm}_i)$ given by Eq.~(\ref{gq}). By expressing the matrix elements of $Z$ as $ Z_{ij}= \delta_{ij} + \int_{-q_i}^{q_i} d \mu\,  g_{ij}(\mu)$, with $g_{ij}(\mu) \equiv  (-1)^{j+1} g(\mu|q_j)$, one finds $Z_{ij} = \xi_{ij}(q_i)$, where $\xi(\lambda)$ is the dressed charge matrix defined through
\begin{equation}
\xi_{ij}(\lambda) -  \frac{1}{2\pi} \sum_k \int_{-\infty}^{\infty} d \mu \, K_{ik}(\lambda,\mu) \xi_{kj}(\mu) = \delta_{ij},
\end{equation} 
so that here the kernel matrix element is $K_{ij} (\lambda, \mu) = K(\lambda,\mu)$ when $-q_j \leq \mu \leq q_j$, otherwise zero.
Now from
\begin{equation}
\hspace{-2cm}   Z_{ij}-\delta_{ij} = \int_{-\infty}^{-q_i} d \lambda \, \frac{\partial \xi_{ij}(\lambda)}{\partial q_i} \Bigg|_{\lambda = q_i} = - \sum_k \int_{-\infty}^{-q_i} d \lambda \, \left[ g_{ki} (\lambda) - g_{ki}(-\lambda) \right] Z_{kj},
\end{equation}
it follows that
\begin{eqnarray}
\hspace{-2cm}  Y_{ij} =  \delta_{ij} + \left( \int_{-\infty}^{-q_i} - \int_{q_i}^{\infty}  \right)  d\mu \, g_{ij}(\mu)  = \delta_{ij} - \sum_k \left[ Z_{ki} - \delta_{ki} \right]  (Z^{-1})_{jk} =  (Z^{-1})_{ji}.
\end{eqnarray}
Hence $ Y = (Z^T)^{-1}$. Putting this into Eqs.~(\ref{dq}) and (\ref{dnt5}), finally gives
\begin{equation} \label{fscsymm}
   \delta E = \frac{2\pi}{L} \sum_j   v_j  \Big[ \Delta_j^+   + \Delta_j^-   \Big] ,
\end{equation}
where
\begin{eqnarray} \label{dims}
\Delta_j^{\pm} = N_j^{\pm} + \frac{1}{2} \left( \sum_k (Z^{-1})_{jk} \frac{ \Delta N_k}{2} \pm  \sum_k Z_{kj}  d_k \right)^2 .
\end{eqnarray}

Obviously, when the bare single-particle dispersion in Eq.~(\ref{e0}) is given by $\varepsilon_0 (\lambda) = \lambda^2$, there is just a single Fermi sea, between $\lambda = -q$ and $\lambda = q$, and Eqs.~(\ref{fscsymm}) and (\ref{dims}) reduce to those for the usual Lieb-Liniger model,
\begin{equation} \label{fsc0}
   \delta E = \frac{2\pi}{L}v  \Big[ \Delta^+   + \Delta^-   \Big] , \qquad \Delta^{\pm} = N^{\pm} + \frac{1}{2} \left( \frac{\Delta N}{2Z(q)}  \pm Z(q) d \right)^2 .
\end{equation}

\subsection{Momentum}

Since the momentum $P=\sum_i \lambda_i = \frac{2\pi}{L} \sum_i n_i$, as is easily seen from Eq.~(\ref{bethe2}), it is trivially obtained as
\begin{eqnarray} \label{fscmom}
\hspace{-1cm}  P  = P_0 + \sum_{i}   \Big[ \,   \tilde{P}_i \,\Delta N_i   - k_i \,  d_i +   \frac{2\pi}{L} \,\left\{ N_i^+ - N_i^- +  d_i \, \Delta N _i \right\} \, \Big],
\end{eqnarray}
for the excited states, where $P_0$ is the ground-state momentum. In the thermodynamic limit $\tilde{P}_i = (q_i^+ + q_i^-)/2$ and $k_i = q_i^+ - q_i^-$. In case of parity symmetry $\tilde{P}_j = 0$ and $k_j = 2q_j$.

\section{Conformal dimensions and correlation functions}

It is now clear that the finite-size corrections in Eqs.~(\ref{eg0L}), (\ref{fscsymm}) and (\ref{fscmom}) can be written on the form
\begin{eqnarray}
E_0 - E_0(L \to \infty) = - \frac{\pi }{6 L}  \sum_{j} c_j | v_j|  + h.o.c. \label{cft1} \\
E - E_0 = \frac{2\pi}{L} \sum_j|v_j|  \left[  \Delta_j^+ +   \Delta_j^- \right] + h.o.c. \label{cft2} \\
P - P_0 = \sum_j \left[  \tilde{P}_j \,\Delta N_j   - k_j \,  d_j \right] +  \frac{2\pi}{L} \sum_j \left[ \Delta_j^+ -  \Delta_j^- \right] + h.o.c. \label{cft3}
\end{eqnarray}
where $h.o.c.$ denotes higher-order corrections. This tells us that the low-energy physics is given by a sum of conformal field theories~\cite{cardy,blote}, where $\Delta_j^{\pm}$ are the scaling dimensions of the scaling operators of the theories and $c_j=1$ the central charges. We will now use this to obtain the large-distance asymptotics of the equal-time correlation functions~\cite{bir}. The field correlation function
\begin{equation}
\langle \Psi(x) \Psi^{\dagger}(0) \rangle = \textrm{Tr} \  \hat{\rho}_{G}  \Psi(x) \Psi^{\dagger}(0),
\end{equation}
for which the total number of added {\it particles} is $\Delta N = \sum_{j} (\Delta N_{2j-1} - \Delta N_{2j})  =1$, should then have the asymptotic form
\begin{equation}
\langle \Psi(x) \Psi^{\dagger}(0) \rangle \sim    \sum_{\mathcal{Q}_{\Psi}} A(\mathcal{Q}_{\Psi} )\,   e^{i \sum_{j} d_{j} k_j x}  \,  x^{-2\sum_j (\Delta_j^+ + \Delta_j^-)},
\end{equation}
where $\mathcal{Q}_{\Psi} = \{ \{\Delta N_{j} \} , \{ d_{j} \} , \{ N_j^{\pm}  \} \, | \, \Delta N =1 \}$ is the set of quantum numbers for the excitations, and $A(\mathcal{Q}_{\Psi})$ the amplitudes.  The leading term is given by
\begin{equation}
\langle \Psi(x) \Psi^{\dagger}(0) \rangle \sim  x^{-\alpha},
\end{equation}
with $\alpha$ the smallest sum of scaling dimensions $2\sum_j (\Delta_j^+ + \Delta_j^-)$ given the constraint $\Delta N = 1$, i.e. $\alpha$ is the smallest of the numbers $\left(\sum_j (Z^{-1})_{jk} \right)^2 /2$. Similarly, density-density correlators are obtained with $\Delta N = 0$,
\begin{equation}
\langle j(x) j(0) \rangle - \langle j(0) \rangle ^2 \sim  \sum_{\mathcal{Q}_{d}} A(\mathcal{Q}_{d} )\,   e^{i \sum_{j} d_{j} k_j x}  \,  x^{-2\sum_j (\Delta_j^+ + \Delta_j^-)},
\end{equation}
where $j(x) = \Psi^{\dagger}(x) \Psi(x)$ and $\mathcal{Q}_{d} = \{ \{\Delta N_{j} \} , \{ d_{j} \} , \{ N_j^{\pm}  \} \, | \, \Delta N = 0 \}$. The leading terms are
\begin{equation}
\langle j(x) j(0) \rangle - \langle j(0) \rangle ^2 \sim  A_1 x^{-2}   + A_2 \cos (d_k k_k x) x^{-\theta},
\end{equation}
where the first term comes from the processes with $d_k = 0$ and one single number $N_k^{\pm}$ equal to one, and the second term is from the two processes with the $d_k = \pm 1$ giving smallest $\theta$, i.e. with $\theta$ the smallest of the numbers $2\left(\sum_j Z_{kj} \right)^2$.

Importantly, the finite-temperature correlation functions for mixed states are obtained by the standard conformal mapping $z(w) = e^{2\pi T w/v_i}$ in the complex plane $z=x-i v_i t$, so that in the formulas above
\begin{equation}  \label{finitetemp}
x^{-2\Delta_j^{\pm}}  \ \to\  \left\{ (\pi T / v_j ) /  \sinh [ \pi T x / v_j] \right\}^{2\Delta_j^{\pm}} .
\end{equation}

\section{Discussion}

We have obtained the finite-size corrections for the energy and momentum of a one-dimensional Bose gas with delta-function interaction (Lieb-Liniger model) when additional conservation laws are present in the density matrix, or the conserved charges are added to the Hamiltonian. The results show that the low-energy physics is described by a sum of conformal field theories each with central charge $c=1$, where each may have its own specific speed of sound. The picture is most clear when the bare dispersion is parity symmetric. In this case we derived the asymptotic behavior of the correlation functions using standard arguments.

The finite temperature mapping (\ref{finitetemp}) provides a possible connection to present studies of quantum non-equilibrium dynamics. In the standard setup, an isolated system in a pure state is time-evolved by, but not an eigenstate of, the usual Schr\"odinger Hamiltonian. The density matrix of a subsystem will however be in a mixed state, presumably approaching the form (\ref{gge}) with a finite effective temperature. It remains to be seen whether the new types of correlation effects in the generalized model studied here may appear in such systems.

It is interesting to note that the generalized one-dimensional Bose gas with a dispersion relation with many Fermi points for the bare particles gives similar types of finite-size effects as in well-studied examples of multicomponent (a.k.a.~nested) Bethe Ansatz solvable models~\cite{ikr}, such as integrable quantum spin chains~\cite{tsvelik,frahm,pokrovskii,fy,fr,z,zkz,zk} and the one-dimensional Hubbard model~\cite{hubbard,fk}, even though the one-dimensional Bose gas only contains a single species of particles.

\section*{Acknowledgments}
We wish to thank Holger Frahm for helpful discussions. EE acknowledges financial support from the Swedish Research Council (Grant No. 621-2011-3942) and from STINT (Grant No. IG2011-2028), and VK from NSF (Grant No. DMS-1205422).


\section*{References}
\begin{thebibliography}{10}


\bibitem{jaynes} Jaynes E T, 1957, Information theory and statistical mechanics, {\it Phys.Rev.} {\bf 106} 620

\bibitem{kinoshita} Kinoshita T, Wenger T and Weiss T D, 2006, A quantum NewtonÕs cradle,
{\it Nature} {\bf 440} 900

\bibitem{rigol} Rigol M, Dunjko V, Yurovsky V and Olshanii M, 2007, Relaxation in a Completely Integrable Many-Body Quantum System: An Ab Initio Study of the Dynamics of the Highly Excited States of 1D Lattice Hard-Core Bosons,
{\it Phys. Rev. Lett.} {\bf 98} 050405

\bibitem{polkovnikov} Polkovnikov A, Sengupta K, Silva A and Vengalattore M, 2011, Nonequilibrium dynamics of closed interacting quantum systems,
{\it Rev. Mod. Phys.} {\bf 83} 863

\bibitem{refs} Calabrese P, Essler F H L and Fagotti M, 2012, Quantum quenches in the transverse field Ising chain: II. Stationary state properties {\it J. Stat. Mech.} (2012) P07022

\bibitem{ce} Caux J-S and Essler F H L, 2013, Time evolution of local observables after quenching to an integrable model, arXiv:1301.3806

\bibitem{ck} Caux J-S and Konik R M, 2012, Constructing the Generalized Gibbs Ensemble after a Quantum Quench, {\it Phys. Rev. Lett.} {\bf 109} 175301

\bibitem{ksci} Kormos M, Shashi A, Chou Y-Z and Imambekov A, 2012, Interaction quenches in the Lieb-Liniger model, arXiv:1204.3889

\bibitem{mc2} Mossel J and Caux J-C, 2012, Exact time evolution of space- and time-dependent correlation functions after an interaction quench in the one-dimensional Bose gas, {\it New J. Phys.} {\bf 14} 075006

\bibitem{bck} Brandino G, Caux J-C, Konik R, 2013, Relaxation dynamics of conserved quantities in a weakly non-integrable one-dimensional Bose gas, arXiv:1301.0308

\bibitem{tsvelik} Tsvelik A M, 1990, Incommensurate phases of quantum	one-dimensional magnetics, {\it Phys. Rev. B} {\bf 42} 779

\bibitem{frahm} Frahm H, 1992, Integrable spin-1/2 XXZ Heisenberg chain with competing interactions, {\it J. Phys. A: Math.Gen.} {\bf 25} 1417

\bibitem{liebliniger} Lieb E H and Liniger W, 1963, Exact Analysis of an Interacting Bose Gas. I. The General Solution and the Ground State,
{\it Phys. Rev.} {\bf 130} 1605

\bibitem{lieb} Lieb E H, 1963, Exact Analysis of an Interacting Bose Gas. II. The Excitation Spectrum,
{\it Phys. Rev.} {\bf 130} 1616

\bibitem{kbi} Korepin V E, Bogoliubov N M and Izergin A G, 1993, {\it Quantum Inverse Scattering Method and Correlation Functions} (Cambridge: Cambridge University Press)

\bibitem{davies} Davies B and Korepin V E, 1989, {\it Higher conservation laws for the quantum non-linear Schr\"odinger equation} (Canberra: Centre for Mathematical Analysis, Australian National University) Report No. CMA-R33-89. Available at arXiv:1109.6604

\bibitem{mossel} Mossel J and Caux J-S, 2012, Generalized TBA and generalized Gibbs, {\it J. Phys. A: Math. Theor.} {\bf 45} 255001

\bibitem{we} Woynarovich F and Eckle H-P, 1987, Finite-size corrections and numerical calculations for long
spin-1/2 Heisenberg chains in the critical region, {\it J. Phys. A: Math. Gen.} {\bf 20} L97

\bibitem{bm} Berkovich A and Murthy G, 1988, Finite-size corrections in the non-linear Schr\"odinger model, {\it J. Phys. A: Math. Gen.} {\bf 21} L395

\bibitem{1979} Korepin V E, 1979, Direct calculation of the S matrix in the massive Thirring model, {\it Theor. Math. Phys.} {\bf 41} 953

\bibitem{bir} Bogoliubov N M, Izergin A G and Reshetikhin N Yu, 1987, Finite-size effects and infrared asymptotics of the correlation functions in two dimensions, {\it J. Phys. A: Math. Gen.} {\bf 20} 5361

\bibitem{wet} Woynarovich F, Eckle H-P and Truong T T, 1989, Non-analytic finite-size corrections in the one-dimensional
Bose gas and Heisenberg chain, {\it J. Phys. A: Math. Gen.} {\bf 22} 4027

\bibitem{w} Woynarovich F, 1989, Finite-size effects in a non-half-filled Hubbard chain, {\it J. Phys. A: Math. Gen.} {\bf 22} 4243

\bibitem{hubbard} Essler F H L, Frahm H, G\"ohmann F, Kl\"umper A and Korepin V E, 2005, {\it The One-Dimensional Hubbard Model} (Cambridge: Cambridge University Press)

\bibitem{zvyagin} Zvyagin A A, 2005, {\it Finite Size Effects in Correlated Electron Models: Exact Results} (London: Imperial College Press) 

\bibitem{cardy} Cardy J L, 1984, Conformal invariance and universality in finite-size scaling, {\it J. Phys. A: Math. Gen.} {\bf 17} L385

\bibitem{blote} Bl\"ote H W, Cardy J L and Nightingale M P, 1986, Conformal invariance, the central charge, and universal finite-size amplitudes at criticality, {\it Phys. Rev. Lett.} {\bf 56} 742

\bibitem{pokrovskii} Pokrovskii S V and Tsvelik A M, 1987, Conformal dimension spectrum for lattice integrable models of magnets, {\it Sov. Phys. JETP} {\bf 66} 1275

\bibitem{ikr} Izergin A G, Korepin V E and Reshetikhin N Yu, 1989, Conformal dimensions in Bethe ansatz solvable models, {\it J. Phys. A: Math. Gen.} {\bf 22} 2615

\bibitem{fy} Frahm H and Yu N-C, 1990, Finite-size effects in the integrable XXZ Heisenberg model with arbitrary spin, {\it J. Phys. A: Math. Gen.} {\bf 23} 2115

\bibitem{fr} Frahm, H and R\"odenbeck C, 1997, Properties of the chiral spin liquid state in generalized spin ladders, {\it J. Phys. A: Math. Gen.} {\bf 30} 4467

\bibitem{z} Zvyagin A A, 2000, Commensurate-incommensurate phase transitions for multichain quantum spin models: exact results, {\it Low Temp. Phys.} {\bf 26} 134

\bibitem{zkz} Zvyagin A A, Kl\"umper A and Zittartz J, 2001, Integrable correlated electron model with next-nearest-neighbour interactions, {\it Eur. Phys. J. B} {\bf 19} 25

\bibitem{zk} Zvyagin A A and Kl\"umper A, 2003, Quantum phase transitions and thermodynamics of quantum antiferromagnets with next-nearest-neighbor couplings, {\it Phys. Rev. B} {\bf 68} 144426

\bibitem{fk} Frahm H and Korepin V E, 1990, Critical exponents for the one-dimensional Hubbard model, {\it Phys. Rev. B} {\bf  42} 10553

\end{thebibliography}
\end{document}